\documentclass[12pt]{iopart}
\usepackage{graphicx}
\usepackage{amstext}

\def\be{\begin{equation}}
\def\ba{\begin{eqnarray}}
\def\a{\alpha}
\def\b{\beta}

\def\d{\delta}

\def\Th{\Theta}

\def\x{\xi}
\def\p{\pi}
\def\r{\rho}

\def\t{\tau}

\def\F{\Phi}

\def\o{\omega}
\def\O{\Omega}
\def\i{\int}

\def\ee#1{\label{#1}\end{equation}}
\def\ea#1{\label{#1}\end{eqnarray}}

\begin{document}
\title[Transition times and phase diffusion]%
      {Statistics of transition times, phase diffusion and
       synchronization in periodically driven bistable systems}
\author{Peter~Talkner$^\ast$,
        {\L}ukasz~Machura$^\ast$,
        Michael~Schindler$^\ast$,
        Peter~H\"anggi$^\ast$\ and
        Jerzy~{\L}uczka$^\ddag$}
\address{$\ast$\ Institut f\"ur Physik, Universit\"at Augsburg, D-86135 Augsburg, Germany}
\address{\ddag\ Institute of Physics,  University of Silesia, 40-007 Katowice, Poland}
\ead{michael.schindler@physik.uni-augsburg.de}
\date{\today}
\begin{abstract}
The statistics of transitions between the metastable states of a
periodically driven bistable Brownian oscillator are investigated on
the basis of a two-state description by means of a master equation
with time-dependent rates. The theoretical results are compared with extensive
numerical simulations of the Langevin equation for a sinusoidal
driving force. Very good agreement is achieved both for the counting
statistics of the number of transitions per period and the residence time
distribution of the process in either state. The counting statistics
corroborate in a consistent way the interpretation of stochastic
resonance as a synchronization phenomenon for a properly defined
generalized Rice phase.
\end{abstract}
\pacs{02.50.Ga, 05.40.-a, 82.20.Uv, 87.16.Xa}
\submitto{\NJP}

\maketitle

\section{Introduction}\label{I}
Time-dependent systems that are in contact with an environment
represent an important class of non-equilibrium systems. In these
systems effects may be observed that cannot occur in thermal
equilibrium, as for example noise-sustained signal amplification by
means of stochastic resonance \cite{GHJM}, or the rectification of
noise and the appearance of directed transport in ratchet-like systems
\cite{ratch}. In the case of stochastic resonance the transitions
between two metastable states appear to be synchronized with an
external, often periodic signal that acts as a forcing on the
considered system \cite{RFNSG,CHLFSG}. This force alone, however, may
be much too small to drive the system from one state to the other.
Responsible for the occurrence of transitions are random forces imposed
by the environment in form of thermal noise. This kind of processes
can conveniently be modeled by means of Langevin equations which are
the equations of motion for the relevant degree (or degrees) of
freedom of the considered system and which comprise the influence of
the environment in the form of dissipative and random forces
\cite{HT}. Unfortunately, neither the Langevin equations nor the
equivalent Fokker-Planck equations for the time evolution of the
system's probability density can be solved analytically for other than
a few rather special situations \cite{JH}. Therefore, most of the prior
investigations of stochastic resonance are based on the numerical
simulation of Langevin equations \cite{GHJM}, or the numerical
solution of Fokker-Planck equations either by means of continued
fractions \cite{JH} or Galerkin-type approximation schemes
\cite{LRH,STH}. Analytical results have been obtained in limiting
cases like for weak driving forces \cite{GHJM}, or weak noise
\cite{LRH}.

Interestingly enough, the same qualitative features that had been
known from numerical investigations also resulted from very simple
discrete models \cite{McW,SJH} that contain only two states
representing the metastable states of the continuous model. The
dynamics of the discrete states is Markovian and therefore governed by
a master equation. The external driving results in a time-dependent
modulation of the transition rates of this master equation. In recent
work \cite{TL} it was shown that such master equations correctly
describe the relevant aspects of the long-time dynamics of
continuously driven systems with metastable states provided that the
intra-well dynamics is fast compared to both the driving and the
noise-activated transitions between the metastable states.

In the present work we will pursue ideas \cite{CHLFSG} suggesting that
a periodically driven bistable system can be characterized in terms of
a conveniently defined phase that may be locked with the phase of the
external force, yielding in this way a more precise notion of
synchronization for such systems. Various possible definitions of
phases for a bistable Brownian oscillator have been compared in
Ref.~\cite{CHLFSG} with the main result that the precise definition
does not matter much in the rate-limited regime where the master
equation applies. Therefore, one will use the definition for which it
is simplest to determine the corresponding phase. If we think of the
system as of a particle moving in a bistable potential under the influence of a
stochastic and a driving force then the Rice
phase provides a convenient definition under the condition that the
particle has a continuous velocity \cite{CHLFSG,R}. This phase counts the
number of crossings of the potential maximum separating the two wells.
The necessary continuity of the velocity is guaranteed if the
hypothetical particle  has a mass and obeys a Newtonian equation of
motion. The continuity of the velocity is lost if the particle is
overdamped and described by an ``Aristotelian equation of motion,''
i.e.\ by a first-order differential equation for the position driven
by a Gaussian white stochastic force. Then the trajectories are known
to be nowhere differentiable and to possess further level crossings in
the close temporal neighborhood of each single one.

We will allow for this jittering character of Brownian trajectories by
setting two different thresholds on the both sides of the potential
maximum and only count alternating crossings. By definition, the
generalized Rice phase increases by $\p$ upon each counted crossing of
either threshold. We will choose the threshold positions in such a way
that an immediate recrossing of the barrier from either threshold is
highly unlikely. Put differently, up to exceptional cases, the
particle will move from the threshold to the adjacent well rather than
to immediately jump back over the barrier to the original well. In the
two-state description of the process we then can identify the phase by
counting the number of transitions between the two states. With each
transition the phase grows by $\p$. In previous work \cite{T} one
of us gave explicit expressions for various quantities characterizing
the alternating point process, comprised of the entrance times into
the states of a Markovian  two-state process with periodic
time-dependent rates. The definition of the Rice phase in that work
was based on the statistics of the entrance times into one particular
state. Some of the statistical details then depend on the choice of
this state. To avoid this ambiguity we here base our definition on the
transitions between both states.

A further quantity that has been introduced in order to characterize
stochastic resonance is the distribution of residence times
\cite{GMS}. It is also based on the statistics of the transition times
and characterizes the duration between two neighboring transitions.

The paper is organized as follows: In Section 2 we introduce the
model, formulate the respective equivalent Langevin and Fokker-Planck
equations and give the formal solutions of the two-state Markov
process with time-dependent rates that follow from the Fokker-Planck
equation. In Section 3 several statistical tools from the theory of
point processes are introduced by means of which the switching
behavior resulting from the master equation can be characterized.
Explicit expressions for various quantitative measures in terms of the
transition rates and solutions of the master equation are derived.
These are the first two moments of the counting statistics of the
transitions from which the growth rate of the phase characterizing
frequency synchronization, its diffusion constant and the Fano factor
\cite{F} follow. Further, the probabilities for $n$ transitions within
one period of the external driving force and an explicit expression
for the residence time distribution in either state are determined. In
Section 4 these quantities are estimated from stochastic simulations
of the Langevin equations and compared with the results of the
two-state theory. The paper ends with a discussion.

\section{Rate description of a Fokker-Planck process}

We consider the archetypical model
of an overdamped Brownian particle moving in a symmetric double well potential
$U(x) = x^4/4 - x^2/2$ under the influence of a time-periodic driving
force $F(t) = A \sin (\O t)$. Throughout the paper we use dimensionless units:
mass-weighted positions are  given in units of the distance between the local
maximum of $U(x)$ and either of the two minima. The unit of
time is chosen as the inverse of
the so-called barrier frequency $\o_0 = \sqrt{-d^2 U(0)/dx^2} = 1$.
The particle's dynamics then can be described by the Langevin equation
\be
\dot{x}(t) = -\frac{d U(x)}{dx} + F(t) + \sqrt{\frac{2}{\b}} \x(t)
\ee{LE}
where $\x(t)$ is Gaussian white noise, with $\langle
\x(t)\rangle =0$ and   $\langle \x(t) \x(s)\rangle = \d (t-s)$. In the
units chosen the inverse noise strength $\b$ equals four times the
barrier height of the static potential divided by the thermal energy
of the environment of the Brownian particle. The equivalent
time-dependent Fokker-Planck equation describes the time-evolution of
the probability density $\r(x,t)$ for finding the process at time $t$
at the position $x$
\be
\frac{\partial}{\partial t} \r(x,t) = \frac{\partial}{\partial x}
\frac{\partial V(x,t)}{\partial x} \r(x,t) + \b^{-1}
\frac{\partial^2}{\partial x^2} \r(x,t)
\ee{FP}
where the time-dependent potential $V(x,t)$ contains the
combined effect of $U(x)$ and of the external time-dependent force
$F(t)$, i.e.,
\be
V(x,t) = U(x) - x F(t).
\ee{V}
We shall restrict ourselves to forces with amplitudes $A$ being small enough
such that $V(x,t)$ has three distinct extrema
$x_1(t)$, $x_2(t)$ and  $x_b(t)$ for any time $t$,
where $x_1(t)$ and $x_2(t)$ are the
positions of the
left and the right minimum, respectively, and $x_b(t)$ that of the
barrier top. The barrier heights as seen from the two minima are
denoted
by $V_\a(t) = V(x_b(t),t)-V(x_\a(t),t)$ for $\a = 1,2$. We further
assume that the particle only rarely crosses this barrier. This will be
the case if the minimal barrier height $V_m = \text{min}_t V_\a(t) $
is still large enough such that $\b V_m > 4.5$. Under this condition
the time-scales of the inter-well and the intra-well motion are widely
separated and the long-time dynamics is governed by a Markovian
two-state
process where the states $\a =1,2$
represent the metastable states of
the continuous process located at the minima at $x_1(t)$ and
$x_2(t)$ \cite{TL}.
The transition rates $r_{\a,\a'}(t)$, i.e.\ the transition
probabilities from the state $\a'$  to the state $\a$ per unit time,
are time-dependent due to the
presence of the external driving. Explicit
expression for the rates are known if the driving frequency is small
compared to the relaxational frequencies $\o_\a(t)$, $\a =1,2$ and
$\o_b(t)$ that are defined by
\ba
\o_\a(t) & = & \left . \sqrt{\frac{\partial^2
V(x,t)}{\partial x^2}
 }\: \right |_{x=x_\a(t)}
  \nonumber \\
\o_b(t) & = & \left . \sqrt{-\frac{\partial^2 V(x,t)}{\partial x^2}}\:
\right |_{x=x_b(t)}.
\ea{oab}
In the limit of high barriers these rates take the form of Kramers'
rates for the instantaneous potential \cite{HTB}, i.e.
\ba
r_{2,1}(t) & = & \frac{\o_1(t) \o_b(t)}{2 \p}\: e^{-\b V_1(t)} \nonumber \\
r_{1,2}(t) & = & \frac{\o_2(t) \o_b(t)}{2 \p}\: e^{-\b V_2(t)}.
\ea{r12}
A more detailed discussion of the validity of these rates in
particular with respect to the necessary time-scale separation is
given in Ref.~\cite{TL}.
More precise rate expressions that contain finite-barrier corrections
are known \cite{HTB} but will not be used here.

In the semi-adiabatic regime
both
the time-dependent rates and the driving force are much slower than
the relaxational frequencies but no relation between the driving
frequency and the rates is imposed~\cite{TL,TNJP}. Then, the
long-time dynamics of the continuous
process $x(t)$
can be reduced to a Markovian two-state process $z(t)$ governed by the
master equation  with the instantaneous rates (\ref{r12})
\ba
\dot{p}_1(t) & = & -r_{2,1}(t)\, p_1(t) + r_{1,2}(t)\, p_2(t) \nonumber \\
\dot{p}_2(t) & = & \mathop{\phantom{-}}r_{2,1}(t)\, p_1(t) - r_{1,2}(t)\, p_2(t)
\ea{me}
where $p_\a(t)$, $\a=1,2$ denotes the probability that $z(t) = \a$.
These equations can be solved with appropriate initial conditions
to yield the following expressions
for the conditional probabilities $p(\a,t\mid \a',s)$ to find the
particle in the metastable state $\a$ at time $t$ provided that it was
at $\a'$ at the earlier time $s\leq t$ :
\ba
p(1,t\mid 1,s) &=& e^{-R(t)+R(s)} + \i_s^t dt'\:e^{-R(t)+R(t')}r_{1,2}(t')
\nonumber \\
p(1,t\mid 2,s) &=&  \i_s^t dt'\:e^{-R(t)+R(t')}r_{1,2}(t')  \nonumber \\
p(2,t\mid 1,s) &=& \i_s^t dt'\:e^{-R(t)+R(t')}r_{2,1}(t')
\nonumber \\
p(2,t\mid 2,s) &=& e^{-R(t)+R(s)} + \i_s^t dt'\:e^{-R(t)+R(t')}r_{2,1}(t')
\ea{cp}
where
\be
R(t) = \i_0^t dt'\: \left (r_{2,1}(t') + r_{1,2}(t') \right ).
\ee{Rt}
If the conditions are shifted to the remote past, i.e.\ for $s\to
-\infty$ they become effectless at finite observation times $t$.
The asymptotic probabilities then read
\ba
p_1^{\text{as}}(t) & = & \i_{-\infty}^t dt' \:e^{-R(t)+R(t')} r_{1,2}(t'),
\nonumber \\
p_2^{\text{as}}(t) & = & \i_{-\infty}^t dt' \:e^{-R(t)+R(t')} r_{2,1}(t').
\ea{pas}
We note that the asymptotic probabilities are periodic
functions of time with the period $T=2\pi/\Omega$ of the driving force, i.e.
\be
p_\a^{\text{as}}(t+T)= p_\a^{\text{as}}(t).
\ee{pT}
Together with the conditional probabilities in eq.~(\ref{cp}) the
asymptotic probabilities $p^\text{as}_\a(t)$ describe the switching
behavior of the process $x(t)$ at long times.

Finally, we introduce the conditional probabilities
$P_\a(t\mid s)$ that the process stays uninterruptedly in the state $\a$
during the time interval $[s,t)$, provided that $z(s)=\a$. They coincide with
the waiting time distribution in the two states
which can be expressed in terms of the
transition rates
\ba
P_1(t\mid s) & = & \exp \left \{ - \i_s^t dt' r_{2,1}(t') \right \}\\
P_2(t\mid s) & = & \exp \left \{ - \i_s^t dt' r_{1,2}(t') \right \}.
\ea{wtp}

\section{Statistics of transition times and measures of synchronization}
The two-state process $z(t)$ is completely determined by those
times at
which it switches from state 2 into state 1, and vice versa from 1 to 2. These
events constitute an alternating point process
$\ldots t_n<t^*_n<t_{n+1}<t^*_{n+1}< \ldots$ consisting of the two point
processes  $t_n$ ($2 \to 1$)
and $t^*_n$ ($1 \to 2$), see Ref.~\cite{CM}.
This alternating point process can
be characterized by a hierarchy of multi-time joint distribution functions. Of
these, we will mainly consider the single- and  two-time distribution functions.
The single-time distribution function $W(t)$ gives
the averaged number of transitions between the two states 1 and 2 in
the time interval $[t, t+dt)$, i.e.,
$W(t)dt = \langle \# \{s\mid  s=t_n\; \text{or}\; s=t^*_n,\;
s \in [t,t+dt) \} \rangle $.
This density of transition times is given
by the sum of the entrance time densities $W_\a(t)$ specifying
the number of entrances into the individual state $\a$
\be
W(t) = W_1(t) + W_2(t).
\ee{W}
The entrance time densities can be expressed in terms of
the transition rates $r_{\a,\a'}(t)$ and
the single-time probabilities $p_\a(t)$, see Ref.~\cite{T}
\be
W_1(t)= r_{1,2}(t) p_2(t) \quad \text{and} \quad
W_2(t)= r_{2,1}(t) p_1(t).
\ee{W12}

The density $W(t)$ determines the average number
of transitions $ N(t,s)$ between the two states
within the time interval $(t,s)$
\be
\langle N(t,s) \rangle = \i_s^t dt' \:W(t').
\ee{N}
In the limiting case described by the asymptotic probability
$p_\a^{\text{as}}(t)$, see eq.~(\ref{pas}), the entrance time
densities $W(t)$ also is a
periodic function of time. Then, the average $\langle
N(t,s)\rangle$ becomes periodic with respect to a joint shift of both
times $t$~and $s$ by a period~$T$
\be
\langle N(t+T,s+T) \rangle = \langle N(t,s) \rangle
\ee{NT}
and, moreover, the number $\langle N(s+n T,s) \rangle$ is independent of the
starting time $s$ and grows proportionally to the number of periods $n$.

There are two two-time transition distribution functions $f^{(2)}(t,s)$
and $Q^{(2)}(t,s)$ that specify
the average product of numbers of transitions
within the infinitesimal interval $[s,s+ds)$  and
the number of transitions in
the later interval  $[t,t+dt)$. These two functions differ
by the behavior of the process between the two times $s$ and
$t$. For the distribution function $f^{(2)}(t,s)$ the process may have
any number of transitions
within the time interval $[s+ds,t)$. It is
given by the sum of the two-time
entrance distribution functions $f_{\a,\a'}(t,s)$
\be
f^{(2)}(t,s) = \sum_{\a,\a'} f_{\a,\a'}(t,s)
\ee{f}
that determine the densities of pairs of entrances into the states $\a$ and
$\a'$ at the prescribed
respective times $t$ and $s$. For $t>s$ they can be expressed by the
transition rates and the conditional probability $p(\bar{\a}, t\mid \a',
s)$, see Ref.~\cite{T}
\be
f_{\a,\a'}(t,s) = r_{\a,\bar{\a}}(t)\: p(\bar{\a}, t\mid \a', s)\:
r_{\a',\bar{\a'}}(s)\: p_{\bar{\a'}}(s)
\ee{faa}
where the bar over a state, $\bar{\a}$,
denotes the alternative of this
state, i.e.~$\bar{1} =2$ and $\bar{2}=1$.
Note that the  conditional probability
$p(\bar{\a}, t\mid \a', s)$ allows for all possible realizations of the
two-state process starting  with $z(s) = \a'$ at time $s$ up to the time $t$
with any number of transitions in between.
For $t<s$ the function
$f_{\a,\a'}(t,s)$ follows from the symmetry $f_{\a,\a'} (t,s) =
f_{\a',\a} (s,t)$.

In the second two-time distribution function $Q^{(2)}(t,s)$
transitions at the times $s$ and $t$ are taken into account only
if no further transitions occur between the prescribed times.
It is again given by a sum of respective two-time
distribution functions for the individual transitions
from $1$ to $2$, and vice versa, and hence reads
\be
Q^{(2)}(t,s) = \sum_\a Q_\a(t,s)
\ee{Q}
where the two-time entrance distribution function
\be
Q_\a(t,s) = r_{\a,\bar{\a}}(t)\: P_{\bar{\a}}(t\mid s)\:
r_{\bar{\a},\a}(s)\: p_\a(s)
\ee{Qaa}
gives the density of an entrance into the state $\bar{\a}$ at time $s$
and an entrance into state $\a$ at time $t$
conditioned upon processes $z(t')= \bar{\a}$, $s<t'<t$ that stay
constant in the time between $s$ and $t$. Therefore, these distribution
functions depend on the waiting time distribution in the
state $\bar{\a}$ as given by eq.~(\ref{wtp}), see also Ref. \cite{T}.

According to the theory of point processes, see Ref. \cite{S},
the second moment of the number of transitions within the time interval
$[s,t)$ results from the two-time distribution function $f^{(2)}(t,s)$ as
\be
\langle N^2(t,s) \rangle = \langle N(t,s) \rangle + 2 \i_s^t dt'
\i_s^{t'} d s' f^{(2)}(t',s').
\ee{N2}
Subtracting the squared average number $\langle N(t,s) \rangle^2$
one obtains the second moment of the number fluctuations
$\langle (\d N(s+\t,s))^2 \rangle$. It is given by an analogous
expression as $\langle N^2(t,s) \rangle$
\be
\langle\d N^2(t,s) \rangle = \langle N(t,s) \rangle + 2 \i_s^t dt'
\i_s^{t'} d s' g(t',s')
\ee{dN2}
where
\be
g(t,s) = f^{(2)}(t,s) - W(t) W(s).
\ee{g}
If the time difference $t-s$ becomes of the order of the maximal
inverse rate, $\max_t r_{1,2}(t)^{-1}$, i.e.\ of the order of the time
scale on which the process becomes
asymptotically periodic, the two-time distribution function factorizes
into the product $W(t) W(s)$ and $g(t,s)$ vanishes for large $t-s$. Consequently, the
double integral on the right hand side of the eq.~(\ref{dN2}) grows
linearly with $t$ in the asymptotic limit $(t-s) \to \infty$. This leads to
a diffusion of the transition number fluctuations,
i.e.\ an asymptotically linear growth
that can be characterized by a diffusion constant:
\be
D(s) = \lim_{\t \to \infty} \frac{\langle \d N^2(s+\t,s) \rangle
}{2 \t}
\ee{D}
which is a periodic function of the initial time $s$ with the period
$T$
of the external driving force. This time-dependence reflects the
non-stationarity
of the underlying process. The diffusion constant $D(s)$ is
proportional to the variance
of the transition number fluctuations during a period of the process
$z(t)$ in the asymptotic periodic limit
\be
D(s) = \frac{\langle \d N^2(T+s,s) \rangle^{\text{as}}}{2 T}
\ee{DN}
where $\langle\:\rangle^{\text{as}}$ indicates the average in the
asymptotic ensemble. In principle, one can shift the window
$[s,T+s)$ where the transition number fluctuations are
determined in such a way that the diffusion constant $D(s)$ attains a
minimum. In the sequel we will not make use of this possibility.
Superimposed on the linear growth of $\langle \d N^2(t,s) \rangle $
there is
a periodic modulation in $t$ with the period $T$.

The comparison of the first moment of the number
of entrances and the second moment of its fluctuations yields the
so-called Fano
factor $F(s)$ \cite{F}; i.e.,
\be
F(s)= \lim_{\t \to \infty} \frac{\langle \d N^2(s+\t,s) \rangle
}{\langle  N(s+\t,s) \rangle} = \frac{\langle \d N^2(s+T,s) \rangle^{\text{as}}
}{\langle  N(s+T,s) \rangle^{\text{as}} }.
\ee{F}
It provides a quantitative measure of the relative number fluctuations and it
assumes the value $F(s)=1$ in the case of a Poisson process. Here, it
is a periodic function of $s$ which attains minima at the same time
$s$ as $D(s)$.

As already mentioned in the introduction, the number of transitions
between the two states can yet be given another interpretation as a
generalized Rice phase
of the random process $x(t)$. At each time instant the process has switched to
another state,
the phase grows by $\p$. The simplest
definition would be the linear relation $\F(t,s) =  \p N(t,s)$
where the phase is set to zero at the initial time $t=s$.
With this definition the phase changes stepwise. A linear
interpolation would lead to a continuously varying phase, but will not
be considered here. Independently from its precise definition,
in the asymptotic periodic regime, both
the average
and the variance of the phase
increase linearly in time, superimposed by a modulation with the period
of the driving force.

A further coarse-graining of the considered periodically driven
process can be obtained by considering the number of transitions
during an interval $[s,t)$.
By
$p_\a(n;t,s)$ we denote the probability that the process assumes the
value $z(s) = \a$ at time $s$ and undergoes $n$ transitions up to the
time instant $t>s$.
Keeping in mind the significance of the waiting time
distributions $P_\a(t\mid s)$, given by eq.~(\ref{cp}),
as the probability of staying uninterruptedly
in the state $\a$  and of the transition rates $r_{\a,\b}(t)$ as the
probability per unit time of a transition from $\b$ to $\a$ one finds
the following explicit expressions for the first few values of $n$
invoking the basic rules of probability theory:
\ba
p_\a(0;t,s) & = & P_\a(t\mid s) p_\a(s)
\nonumber \\
p_\a(1;t,s) & = & \i_s^t dt_1 P_{\bar{\a}}(t\mid t_1)
r_{\bar{\a},\a}(t_1) P_\a(t_1\mid s) p_\a(s) \\
p_\a(2;t,s) &= & \i_s^t dt_2 \i_s^{t_2} dt_1
P_\a(t\mid t_2) r_{\a,\bar{\a}}(t_2) P_{\bar{\a}}(t_2\mid t_1)
r_{\bar{\a},\a}(t_1) P_\a(t_1\mid s) p_\a(s) \nonumber
\ea{Pan}
where the states $\a$ and $\bar{\a}$ are opposite to each other.
The probabilities with values of $n>0$ can be determined
recursively from the following set of differential equations:
\ba
\frac{\partial}{\partial t} p_\a(n+1;t,s) & = & -r_{\a_{n+1},\a_n}(t)
p_\a(n+1;t,s) + r_{\a_n, \a_{n-1}}(t) p_\a(n;t,s) \nonumber \\
p_\a(n+1;s,s) & = & 0
\ea{dPan}
where
\be
\a_n = \left \{
\begin{array}{ll}
\a \quad& \text{for}\; n\; \text{even}\\
\bar{\a} & \text{for}\; n\; \text{odd}.
\end{array}
\right .
\ee{an}
The hierarchy starts at $n=0$ with $p_\a(0;t,s)$ as defined in
eq.~(\ref{Pan}).
Accordingly, the probability $P(n;t,s)$ of $n$ transitions within
the time interval $[s,t)$ consists in the sum of the individual probabilities
  $p_\a(n,t,s)$
\be
P(n;t,s) = p_1(n;t,s) + p_2(n;t,s).
\ee{Pnts}

Finally, we like to emphasize that with the information at hand also the
dis\-tri\-bu\-tions of residence times can be determined. The residence
times, say of the state $\a$, are defined as the duration of the
subsequent episodes in which the process
dwells in state $\a$ without interruption. For a non-stationary process
these times must not be confused with the life times of this
state. Rather, the distribution of the residence times $R_\a(\t)$
is
the life-time distribution in the state $\a$ averaged
over the starting time with the properly
normalized entrance density of the considered state \cite{lc,cfj}, i.e.,
\be
R_\a(\t) =\frac{\i_0^Tds\: P_\a(s+\t\mid s) r_{\a,\bar{\a}}(s)
  p_{\bar{\a}}^{\text{as}}(s)}{\i_0^T ds r_{\a,\bar{\a}}(s)
  p_{\bar{\a}}^{\text{as}}(s)}.
\ee{Ra}
Here we assumed that the process has reached its asymptotic periodic
behavior and expressed the entrance time distribution according to
eq.~(\ref{W12}).
The denominator guarantees for the proper normalization of the
residence time distribution.

\section{Simulations}

We performed numerical simulations of the Langevin equations by means
of an Euler algorithm with a step size of $10^{-3}$ and with a random
number generator described in Ref.~\cite{RNG}. The comparison with
different random number generators~\cite{RNG2} did not produce any
sensible differences.
We always started the simulations
at the left minimum $x=-1$ at time $t=0$ and determined the first
switching time $t_1$ as the first instant of time at which the value
$x=1/2$ was reached. The switching time $t_1^*$ to the left well was
defined as the first time larger than $t_1$ at which the opposite threshold at
$x=-1/2$ was reached. For $t_2$ we waited until the positive threshold
at $x=1/2$ was again crossed, and so on. In this
way a series of switching times
$t_1,t^*_1,t_1,t^*_2,t_3, \ldots $ was generated. We stopped the
simulations after the time at which either $10^4$, or
at least $10^3$ periods of the
driving force and at
least 500 transitions to the right hand side of the potential
had occurred. The left
and right thresholds were taken in such a way that any multiple
counting due to the inherent irregularities of the Langevin trajectories was
excluded.
Because of the time-scale separation between the inter- and intra-well
dynamics the precise value of the thresholds at $x=\pm 1/2$
is immaterial as long as
they stay in a finite distance from the top of the barrier. Then a
fast return of a trajectory to the previously occupied
metastable state
after a crossing of the threshold
can safely be excluded.

\begin{figure}
\mathindent=0pt
\begin{minipage}[t]{0.48\textwidth}
\centering
\includegraphics{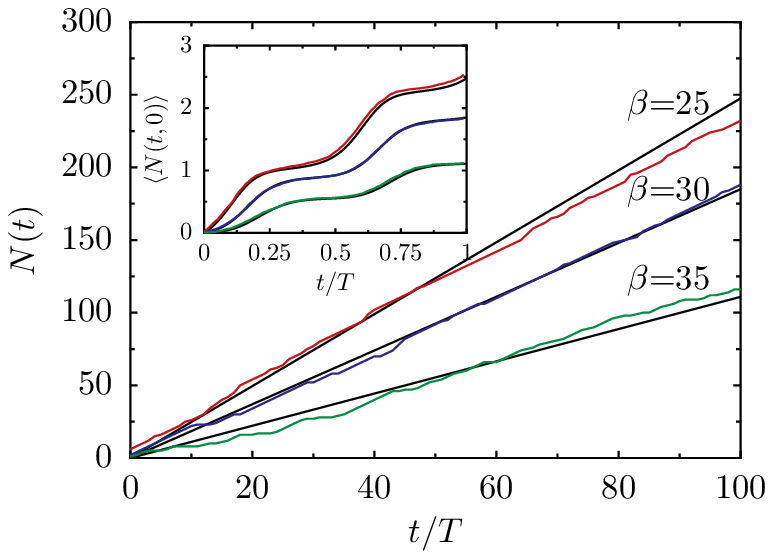}
\caption{The number of transitions $N(t)$ accumulated from $0$ up to
  time $t$ as a function of $t$ for simulations with the driving
  strength $A=0.1$, driving period
  $\O = 10^{-3}$, and inverse
  temperatures $\b =25$ (red), $30$ (blue) and $35$ (green) is
  depicted  together with  the average behavior (black straight lines)
  resulting from
  the two-state theory, see eq.~(\ref{N}). Note that the observed deviations
  apparently are smallest for the middle inverse temperature $\b =30$.
  The mean value $\langle N(t) \rangle$ obtained from
  the simulations as the average over all available periods is
  compared
  with the theoretical prediction in eq.~(\ref{N}).}
\label{f1}
\end{minipage}
\hfill
\begin{minipage}[t]{0.48\textwidth}
\centering
\includegraphics{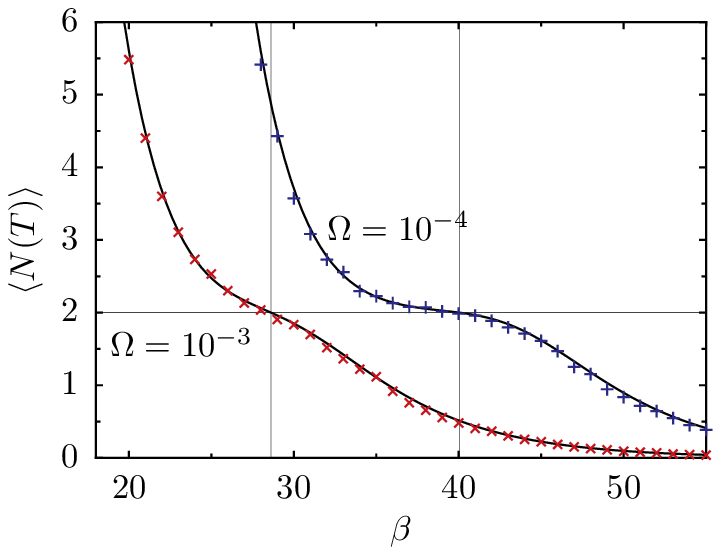}
\caption{The average number of transitions $\langle N(T) \rangle
  \equiv \langle N(T,0) \rangle $ occurring in
  one period is shown as a function of the inverse temperature for the
  two driving frequencies $\O = 10^{-3}$ (red $\times$) and $10^{-4}$
  (blue $+$) and the driving strength $A=0.1$.
The symbols representing the results of the simulations
  nicely fall onto the respective theoretical curves from equation~(15) shown as black
  lines.
  At the temperatures (thin black vertical lines)
  where  $\langle N(T) \rangle $ assume
  the value 2 indicated by the thin black horizontal line,
  the dynamics is optimally
  synchronized with the external driving force. For the smaller
  frequency the average number of transitions flattens around this
  optimal temperature value indicating a tighter locking of the transitions
  with the external driving force. }
\label{f2}
\end{minipage}
\end{figure}

In the simulations the amplitude of the force was always $A=0.1$. For
the frequency we considered the two values $\O= 10^{-3}$ and $\O=
10^{-4}$ and the inverse temperature $\b$ was varied from $20$ to $55$
in integer steps. Note again that the barrier height of the reference
potential $U(x)$ is $\b/4$ in units of the thermal energy.
In Fig.~\ref{f1} the numbers of transitions $N(t)\equiv N(t,0)
= \sum \left (\Th(t-t_i) + \Th(t-t^*_i)\right )$
up to a time $t$ as a function of $t$ are depicted for different
inverse temperatures together with averages estimated from the full
trajectory and those averages following from the two-state model,
see eq.~(\ref{N}). The average growth behavior is determined by the
average number of transitions per period in the asymptotic state,
$\langle N(T+s,s) \rangle^{\text{as}}$, which is independent of
$s$.
In Fig.~\ref{f2} the estimated  value of this number  is compared with
the
two-state model prediction
as a function of $\b$.
The agreement between theory and simulations is excellent even for the
relatively high temperature values and corresponding low transition
barriers for the values of $\b < 25$. The average number of
transitions monotonically decreases with falling temperature. At the
temperature where  $\langle N(T+s,s)\rangle^{\text{as}}=2$ the system
is optimally synchronized with the driving force in the sense that the
Rice phase increases on average by $2 \p$ per period of the driving force. This
optimal temperature depends on the driving frequency and becomes
lowered
for slower
driving. In the vicinity of the optimal temperature the decrease of
$\langle N(T+s,s)\rangle^{\text{as}}$ with inverse temperature is
smallest. The emerging flat region resembles the locking of a driven
nonlinear oscillator and becomes more pronounced for smaller driving
frequencies.

\begin{figure}
\mathindent=0pt
\begin{minipage}[t]{0.48\textwidth}
\includegraphics{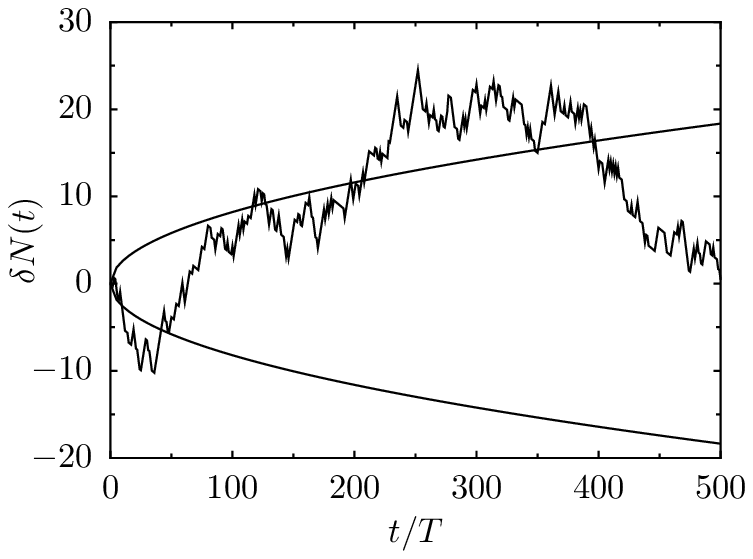}
\caption{The fluctuations of the number of transitions
$\d N(t) $ for $A=0.1$, $\O = 10^{-3}$ and $\b =35$ (jagged line)
appears as diffusive. The smooth curve depicts the root behavior of the diffusion law
$\langle \d N^2(t) \rangle = 2 D(0) t$ with the diffusion constant $D(0) =
\langle \d N^2 (T,0) \rangle /(2 T) = 5.36 \times 10^{-5}$ resulting from the
two-state model, see eq.~(\ref{DN}).}
\label{f3}
\end{minipage}
\hfill
\begin{minipage}[t]{0.48\textwidth}
\includegraphics{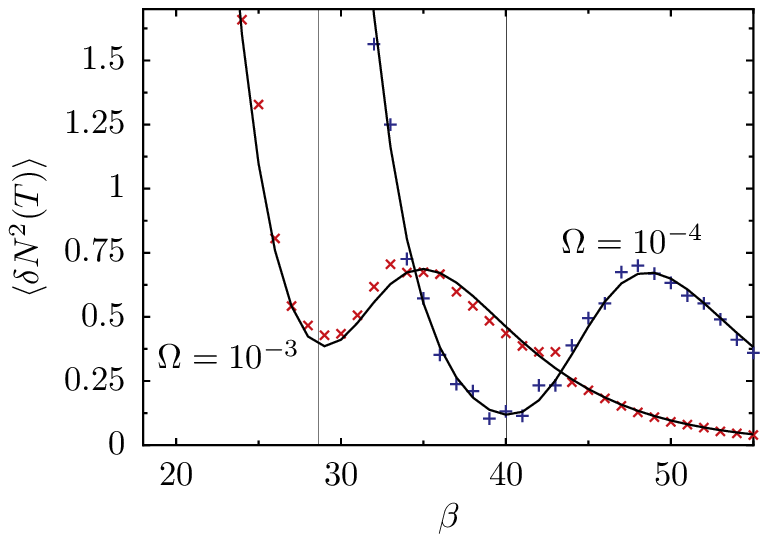}
\caption{The variance of the number of transitions
$\langle \d N^2(T) \rangle \equiv \langle \d N^2(T,0) \rangle $
in one period is shown as a function of the inverse temperature. The
symbols are the same as in Fig.~\ref{f2}, the solid lines display equation (22). Note that $\langle \d N^2(T)
\rangle$ is proportional to the diffusion constant of the Rice phase.
The vertical thin
black lines indicate the optimal inverse temperatures as found from
the synchronization of the averaged phase (see  Fig.~\ref{f2}). They
coincide remarkably well with the positions of the minima of $\langle
\d N^2(T) \rangle$. So we find that at the optimal temperature also
the fluctuations of the number of transitions and consequently of the
generalized Rice phase are suppressed.}
\label{f4}
\end{minipage}
\end{figure}

The fluctuations of the
number of transitions $\d N(t) \equiv N(t,0) - \langle N(t,0)\rangle $
for times up to $t=500 T$
in a single simulation together with
the theoretical average following from eq.~(\ref{dN2}) are
shown in Fig.~\ref{f3}.
The average behavior is characterized by the number fluctuations per
period  $\d N^2(T,0)$. This quantity was estimated from the simulations
as a function of the inverse
temperature. In Fig.~\ref{f4} we compare
the prediction of the two-state model, see eq.~(\ref{dN2}) with numerics.
These number fluctuations exhibit a local minimum very close to
the optimal temperature
where two transitions per period occur on average, see
Figs.~\ref{f2} and \ref{f4}.
This minimum is
more pronounced for the lower driving frequency.
The fluctuations of the Rice phase also assume a minimum at this
optimal temperature. This means that the phase diffusion is minimal at
this temperature.

\begin{figure}
\mathindent=0pt
\begin{minipage}[c]{0.48\textwidth}%
\includegraphics{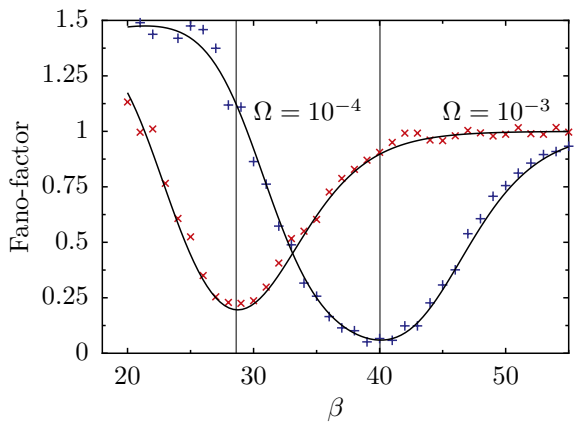}%
\end{minipage}\hfill
\begin{minipage}[c]{0.48\textwidth}%
\caption{As a relative measure of number and phase fluctuations, the
  Fano factor $F=F(0)$ from equation~(26) is shown as a function of the inverse
  temperature. Symbols are the same as in Fig.~\ref{f2}. The minimum
  positions of the Fano factor again nicely coincide with the optimal
  temperature values indicated by the thin black lines. With
  decreasing frequency these minima move to lower temperatures and
  become broader and deeper. For sufficiently low temperatures the
  transitions become very rare and almost Poissonian leading to the
  asymptotic low temperature limit $F=1$.
}%
\label{f5}%
\end{minipage}%
\end{figure}

For the Fano factor $F(0)$ we obtain an absolute minimum near the optimal
temperature, see Fig.~\ref{f5}.
For higher temperatures the Fano factor may become larger
than one, whereas it approaches the Poissonian value $F=1$ for low
temperatures because then, transitions become very rare and
independent from each other.
Also here, theory and simulations agree indeed very well.

\begin{figure}
\mathindent=0pt
\begin{minipage}{0.48\textwidth}
\includegraphics{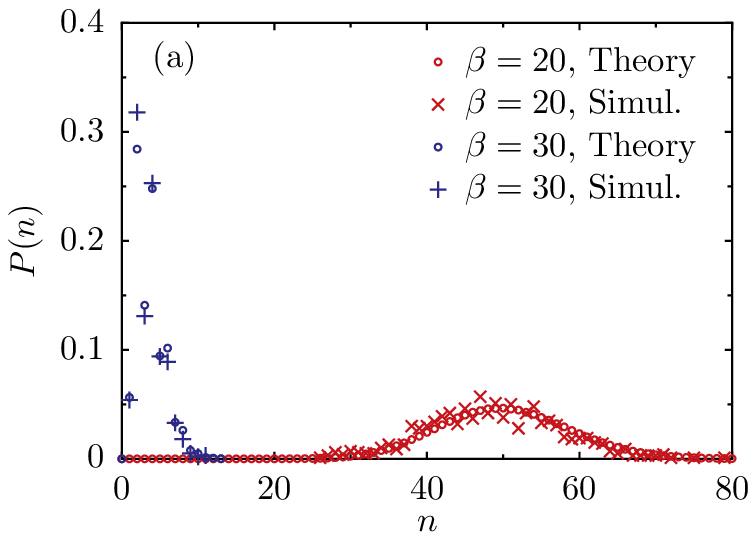}
\end{minipage}
\hfill
\begin{minipage}{0.48\textwidth}
\includegraphics{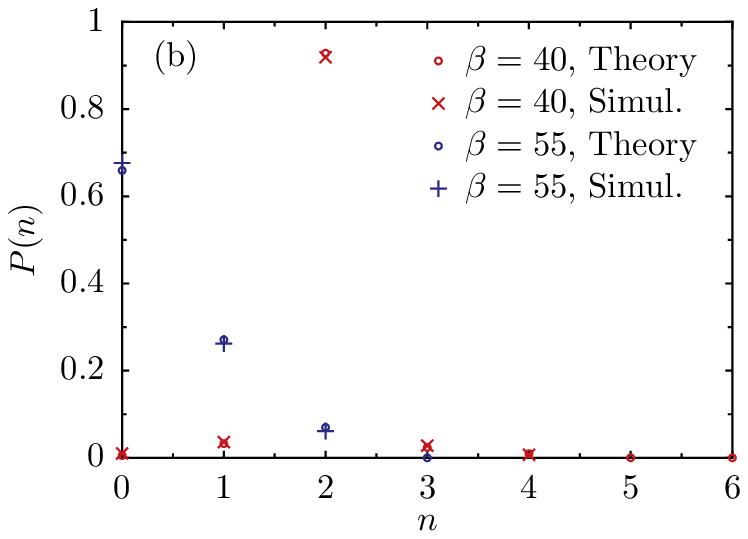}
\end{minipage}
\caption{The probabilities $P(n) = P(n;T,0)$
  are shown for the driving the driving strength $A=0.1$ and the
  driving frequency
  $\O= 10^{-4}$; for two temperatures that are higher than the optimal
  one corresponding to $\b \approx 40$ in panel (a) and and for the
  optimal and a lower temperature in panel (b). As one expects for a
  good synchronization of the system with the driving force at the
  optimal temperature one finds two transitions per period with about
  $90\%$ of the total weight. For high temperatures the distribution
  becomes rather broad with a maximum at some large number of
  transitions. On the contrary, for low temperatures the probability
  is largest at $n=0$ and decreases with $n$. The agreement of the
  two-state theory resulting from equation~(30) with the simulations is remarkably good. }
\label{f6}
\end{figure}

Next, we consider the probabilities $P(n) = P(n;T,0)$ for finding $N(T,0)=n$
transitions per period in the asymptotic, periodic limit. For this purpose we
count the number $k$ of transitions within each period $[nT,(n+1)T)$
and determine their relative frequency occurring in
a simulation.
A comparison with the prediction of the two-state model determined
from the numerical integration of eq.~(\ref{Pan}) and from
eq.~(\ref{Pnts})
is collected in Fig.~\ref{f6} for different
temperature values. The agreement between simulations and theory is
within the expected statistical accuracy. For large temperatures there is
a rather broad distribution of $n$-values around a most probable value $n^*$.
With decreasing temperature, the most probable value moves to smaller
numbers whereby the width of the distribution shrinks. Once $n^*=2$
the probability
$P(2)$ further increases at the cost of the other $k$ values
with decreasing temperature until $k=0$
gains the full weight in the limit of low temperatures.

\begin{figure}
\mathindent=0pt
\begin{minipage}[t]{0.48\textwidth}
\includegraphics{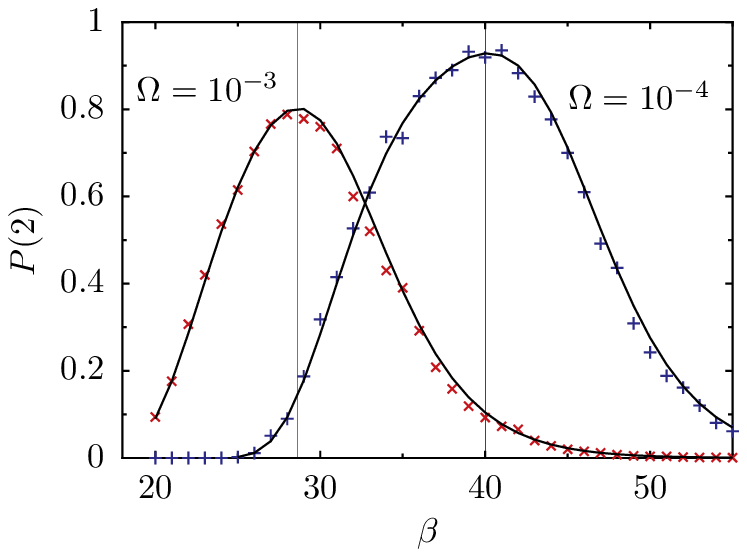}
\caption{The probability $P(2)=P(n=2;T,0)$ for two transitions within one period is
  shown as a function of the inverse temperature $\b$. Results from
  the simulations are depicted by red ($\times$) and blue ($+$)
  crosses for two frequencies. They agree well with the respective
  theoretical predictions of the two-state theory resulting from equation~(30)
  (solid lines). The probabilities
  show the typical stochastic resonance behavior with a maximum
  very close to the optimal temperature. As for the minimum of the
  Fano factor, this maximum is more pronounced at the lower driving
  frequency.
}
\label{f7}
\end{minipage}\hfill
\begin{minipage}[t]{0.48\textwidth}
\includegraphics{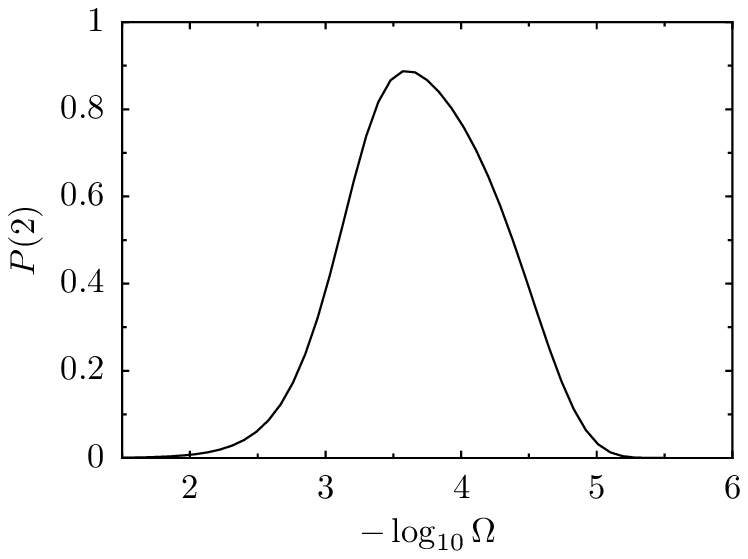}
\caption{The probability $P(2)=P(n=2;T,0)$ (see equation~(30)) for two transitions within one period is
  shown as a function of the frequency at the inverse temperature $\b=
  35$. It also shows a resonance like behavior with a maximum at an
  optimal frequency.}
\label{f8}
\end{minipage}
\end{figure}

The degree of synchronization of the continuous bistable dynamics with
the external driving force can be characterized by the value of the
probability $P(2)$. As a function of the inverse temperature $\b$ it
has a maximum close to the corresponding optimal temperature, see
Fig.~\ref{f7}. For the longer driving period the maximum is at
a lower temperature and its value is higher. Fig.~\ref{f8}
depicts $P(2)$  as a function of the frequency at a fixed
temperature.
It has a maximum at some optimal value of the frequency.

\begin{figure}
\mathindent=0pt
\begin{minipage}{0.48\textwidth}
\includegraphics{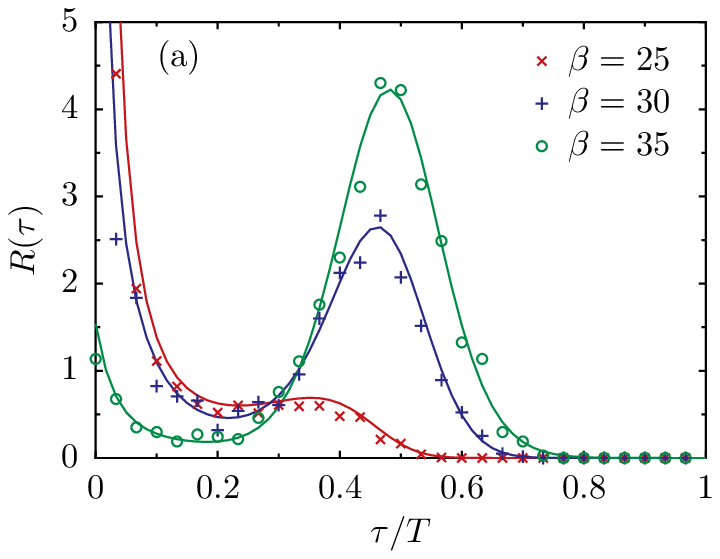}
\end{minipage}
\hfill
\begin{minipage}{0.48\textwidth}
\includegraphics{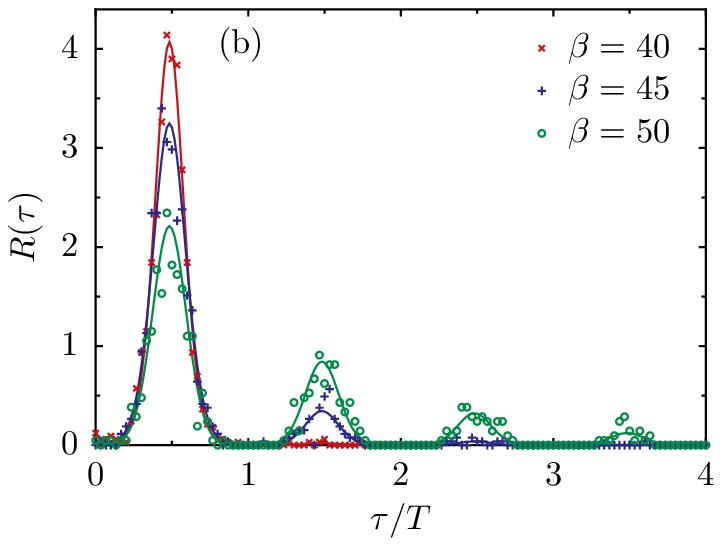}
\end{minipage}
\caption{Residence time distributions for either of the two states are
shown for $\O =10^{-4}$ and different values of the temperature.
Theoretical results obtained from (eq.~\ref{Ra}) are displayed as solid
lines, estimates from the simulations as symbols. In panel (a) $R(\t)$
is shown for higher
temperatures and in panel (b) for the optimal and lower than optimal
temperatures. The peak structure and its temperature dependence is in
complete agreement with previous findings, see Ref.~\cite{GHJM}.}
\label{f9}
\end{figure}

Finally, we come to the residence time distributions which can
be estimated from histograms of the simulated data.
In the
Fig.~\ref{f9} these histograms are compared with the results
of the two-state model. Theory and simulations are in good agreement
and show a transition from a multi-modal distribution with peaks at odd
multiples of half the period  to a mono-modal distribution at zero as
one would expect. At stochastic resonance taking place at the inverse
temperature $\b\approx40$ the distribution is also
mono-modal with its maximum close to half the period.

\section{Discussion and conclusions}
In this work we investigated the statistics of transitions between the
two states of a periodically driven overdamped bistable Brownian
oscillator. We compared results from extensive numerical simulations
with theoretical predictions of a discrete Markovian two-state model
with time-dependent rates. For the studied slow driving frequencies the
rates can be taken as the adiabatic, or frozen, rates. Although the
minimal barriers separating the states from each other may become
rather small measured in units of the thermal energy ($\b
V_{\text{min}} \approx 4$ for the lowest inverse temperature $\b=20$) we
did not take into account finite barrier corrections for the rates and
still obtained very good agreement for
all considered quantities even at the highest chosen temperatures.
We considered different quantities from the counting statistics of
transitions such as the average and variance, as well as the
probability of a given number of transitions in one period. These
quantities are directly related to the average Rice phase, its
diffusion and the probability of a given increase of this phase. We
note that, apart from
the average phase, these quantities depend on the
position of the chosen period for which we took $s=0$ as the starting time.
A different, more sophisticated, way of analyzing the counting statistics
would have been to
position the averaging window of the length of a period
such that the variance of the number of transitions
is minimal. But already with the present simpler method we could detect
significant
signatures of synchronization in a noisy nonlinear system such as the
\emph{locking of frequency} and \emph{phase}
which both occur at basically the
same parameter values.

The novel analytical expression for the residence time distribution
derived for the two-state model also agrees very well with the simulation
results.
\ack
The authors thank Dr. Gerhard Schmid for valuable
discussions. Financial support was provided by the Deutsche
Forschungsgemeinschaft, SFB 438, KBN-DAAD, the Graduiertenkolleg 283, and the
ESF program STOCHDYN.

\end{document}